\documentclass[twocolumn,showpacs,preprintnumbers,prb,aps,psfig]{revtex4}
\usepackage{graphicx}
\usepackage{epstopdf}
\usepackage{float}
\usepackage{color}
\usepackage{amsmath}

\begin{document}

\title{The suppression of electron correlations in the collapsed tetragonal phase of CaFe$_2$As$_2$ under ambient pressure demonstrated  by $^{75}$As NMR-NQR measurements} 
\author{Y. Furukawa, B. Roy, S. Ran, S. L. Bud'ko, and P. C. Canfield}
\affiliation{Ames Laboratory, U.S. DOE, and Department of Physics and Astronomy, Iowa State University, Ames, Iowa 50011, USA}

\date{\today}

\begin{abstract} 
     The static and the dynamic spin correlations in the low temperature collapsed tetragonal and the high temperature tetragonal phase in CaFe$_2$As$_2$ have been investigated by $^{75}$As nuclear magnetic resonance (NMR) and nuclear quadrupole resonance (NQR) measurements.
   Through the temperature ($T$) dependence of the nuclear spin lattice relaxation rates (1/$T_1$) and the Knight shifts, although stripe-type antiferromagnetic (AFM)  spin correlations are realized in the high temperature tetragonal phase, no trace of the AFM spin correlations can  be found in the non-superconducting, low temperature, collapsed tetragonal (c${\cal T}$) phase.  
    Given that there is no magnetic broadening in $^{75}$As NMR spectra, together with the $T$-independent behavior of magnetic susceptibility $\chi$ and the $T$ dependence of 1/$T_1T\chi$, we conclude that Fe spin correlations are completely quenched statically  and dynamically in the non-superconducting  c${\cal T}$ phase in CaFe$_2$As$_2$.

\end{abstract}

\maketitle

      The frequent incidence of magnetism and particularly the role played by antiferromagnetic (AFM) spin correlations on superconductivity and on normal state properties has received wide interest in the study of unconventional superconductors such as high $T_{\rm c}$ cuprates and iron pnictides.\cite{Walstedt, Uemura2010, Johnston2010, Canfield2010, Stewart2011}  
    Among the iron pnictide superconductors, $A$Fe$_2$As$_2$ ($A$ = Ca, Ba, and Sr), known as "122" compounds with a ThCr$_2$Si$_2$-type structure at room temperature, has been one of the most widely studied systems in the recent years.\cite{Johnston2010,Canfield2010,Stewart2011}   
    Application of pressure and carrier doping are considered to play an important role in the suppression of the AFM ordering and the appearance of high temperature superconducting (SC) phase.
    These tuning parameters produce the well-known phase diagram of the Fe-based superconductors: an AFM ordering temperature $T_{\rm N}$  is suppressed continuously with substitution or pressure application, and  an SC state emerges with the transition temperature $T_{\rm c}$  varying as function of the tuning parameters.\cite{Johnston2010,Canfield2010,Stewart2011}

     CaFe$_2$As$_2$ is one such  compound exhibiting an AFM ordering of the Fe moments at $T_{\rm N}$ = 170 K  with a concomitant structural phase transition to a low temperature (LT) orthorhombic  (${\cal O}$) phase. \cite{Ni2008, Goldman2008, Canfield2009}   
  Under ambient pressure, substitutions of Fe by Co, Ni and others induce superconductivity in CaFe$_2$As$_2$ with $T_{\rm c}$  up to $\sim$ 15 K. \cite{Canfield2009,Kumar2009_1,Kumar2009_2,Ran2012}   
     Under a pressure of just a few kilobars, the LT AFM  ${\cal O}$ phase was found to translate  to a non-magnetic collapsed tetragonal (c${\cal T}$) phase.\cite{Canfield2009,Torilachvili2008,Lee,Yu}
    The c${\cal T}$ phase in CaFe$_2$As$_2$ is characterized by a $\sim$10 $\%$ reduction in the tetragonal $c$ lattice constant, from the value in the high temperature (HT) tetragonal (${\cal T}$) phase, along with the absence of  AFM ordering in LT ${\cal O}$ phase.\cite{Kreyssig2008,Goldman2009_2, Ran2011}
     In the case of the presence of a nonhydrostatic pressure component,  an SC phase can be detected,\cite{Canfield2009,Yu,Torilachvili2008} which is considered to originate from a non-collapsed tetragonal phase being stabilized as part of a mixture of several crystallographic phases in CaFe$_2$As$_2$ sample at low temperatures.\cite{Prokes2010, BaekCaFe2As2_P}

    Recently the c${\cal T}$ phase was induced in CaFe$_2$As$_2$ under ambient pressure by changing the heat treatment conditions  which control strains inside crystal grown out of excess FeAs due to formation of nanoscale precitipates. 
    According to Ran ${\it et ~ al.}$\cite{Ran2012,Ran2011} CaFe$_2$As$_2$ annealed at 400~$^{\circ}$C for 24 hours undergoes a phase transition from the HT  ${\cal T}$ paramagnetic state to the LT ${\cal O}$ AFM state at $T_{\rm N}$  $\sim$ 170 K,
similar to the CaFe$_2$As$_2$ crystals grown with Sn flux. 
      On the other hand, CaFe$_2$As$_2$ grown out of excess FeAs,  quenched from 960~$^{\circ}$C to room temperature, exhibits a transition to the c${\cal T}$ non-magnetic phase below $T_{\rm s}$ $\sim$ 95 K. 
       This  opens up opportunities for detailed investigations of the c${\cal T}$ phase in CaFe$_2$As$_2$ under  ambient pressure.
       Quite recently detailed inelastic neutron scattering (INS)\cite{Soh2013} and angle-resolved photoemission spectroscopy (ARPES)\cite{Dhaka2014} studies on the quenched CaFe$_2$As$_2$ crystal under  ambient pressure have been carried out.
       The absence of magnetic fluctuations in the non-superconducting c${\cal T}$ phase was evidenced by the INS study, which indicates that spin fluctuations are a necessary ingredient for unconventional SC in the iron pnictides.
     However, INS measurements, probing mainly high energy spin dynamics (order of K), could not exclude the presence of magnetic fluctuations completely in very low energy regions such as mK order.

     Nuclear magnetic resonance (NMR) and nuclear quadrupole resonance (NQR) can detect low energy spin dynamics through nuclear spin lattice relaxation rate (1/$T_1$) measurements. 
     Kawasaki ${\it et~al}$.\cite{Kawasaki2010,Kawasaki2011} reported the $T$-independent behavior of 1/$T_1T$ on $^{75}$As-NQR in the c${\cal T}$ phase in CaFe$_2$As$_2$ under pressure of 10.8 kbar, demonstrating the absence of the superconducting state in the c${\cal T}$ phase. 
    Ma ${\it et~al}$. \cite{Ma2013} have carried out $^{75}$As NMR in  Pr-doped CaFe$_2$As$_2$
and found a large increase in a nuclear quadrupole frequency $\nu_{\rm Q}$ in the LT c${\cal T}$ phase, similar to the case of $^{75}$As NMR in the quenched CaFe$_2$As$_2$.\cite{Ran2011} 
   They have also reported the $T$ dependence of 1/$T_1$ in the c${\cal T}$ phase showing a broad peak at ${\sim}$25 K, which is attributed to originate from Pr$^{3+}$ spin dynamics, masking out intrinsic properties of spin correlations of Fe spins. 
    Given these results, a detailed study of static and dynamic spin correlations in the c${\cal T}$  phase in non rare-earth bearing CaFe$_2$As$_2$ measured under ambient pressure is intriguing and important, and also would provide some clues about the origin of recent-discovered SC in the c${\cal T}$ phase of (Ca$_{1-x}$Sr$_x$)Fe$_2$As$_2$ with $T_{\rm c}$ $\sim$ 22 K in Ref.~\onlinecite{Jeffries2012} and (Ca$_{1-x}$R$_x$)Fe$_2$As$_2$ (R = Pr, Nd) with $T_{\rm c}$ $>$ 45 K in Ref.~\onlinecite{Saha2012}, as well as for the observation of SC in other carrier-doped CaFe$_2$As$_2$.\cite{Lv2011,Danura2011,Kudo2013}

      In this paper, we report  $^{75}$As NMR measurements to investigate electronic and magnetic properties of the c${\cal T}$ phase in CaFe$_2$As$_2$.   
   From the $T$ dependence of the 1/$T_1$, stripe-type AFM spin correlations are realized in the HT ${\cal T}$ phase. 
   On the other hand, no trace of the AFM spin correlations can be found in the non-superconducting LT c${\cal T}$ phase, demonstrating a quenching of Fe moments in the LT c${\cal T}$ phase from a microscopic point of view.
   These results are consistent with the recently reported INS\cite{Soh2013} and ARPES\cite{Dhaka2014} results.

    
   The single crystals of CaFe$_2$As$_2$  used in this study were grown out of an FeAs flux,\cite{Ran2011,Ran2012} using conventional high temperature growth techniques.\cite{Canfield_book, Canfield_1992} 
   A single crystal, referred to as ''as-grown'',  was quenched from 960~$^{\circ}$C to room temperature.
   The as-grown crystal shows a HT ${\cal T}$ - LT c${\cal T}$ phase transition at $T_{\rm s}$ $\sim$ 96 K.
   For comparison,  we also carried out NMR measurements on the other CaFe$_2$As$_2$ crystals, referred to as  "annealed", which were annealed at 400~$^{\circ}$C for 24 hours and then quenched to room temperature. 
    The annealed CaFe$_2$As$_2$ undergoes a phase transition from the HT ${\cal T}$ paramagnetic state to a LT ${\cal O}$ AFM state at $T_{\rm N}$ $\sim$ 170 K. 
    Details of the growth, annealing and quenching procedures are reported in Refs~\onlinecite{Ran2012} and \onlinecite{Ran2011}.

    NMR and NQR measurements were carried out on $^{75}$As  (\textit{I} = 3/2, $\gamma/2\pi$ = 7.2919 MHz/T, $Q$ =  0.29 Barns)  by using a homemade, phase-coherent, spin-echo pulse spectrometer.  
   The $^{75}$As-NMR spectra were obtained by sweeping the magnetic field $H$ at a fixed frequency $f$ = 51 MHz, while $^{75}$As-NQR spectrum in zero field was measured in steps of frequency by  measuring the intensity of the Hahn spin-echo.  
   The magnetic field was applied parallel to either the crystal $c$-axis or the $ab$-plane. 
    The $^{75}$As 1/$T_{\rm 1}$ was measured with a saturation recovery method. 
   The $1/T_1$ at each $T$ was determined by fitting the nuclear magnetization $M$ versus time $t$  using the exponential functions $1-M(t)/M(\infty) = 0.1 e^ {-t/T_{1}} +0.9e^ {-6t/T_{1}}$ for $^{75}$As NMR,  and  $1-M(t)/M(\infty) = e^ {-3t/T_{1}}$ for $^{75}$As NQR,     where $M(t)$ and $M(\infty)$ are the nuclear magnetization at time $t$ after the saturation and the equilibrium nuclear magnetization at $t$ $\rightarrow$ $\infty$, respectively. 
   Preliminary results of the $^{75}$As-NMR spectra for the as-grown and annealed CaFe$_2$As$_2$ samples have been reported previously.\cite{Ran2011}

\begin{figure}[tb]
\includegraphics[width=8cm]{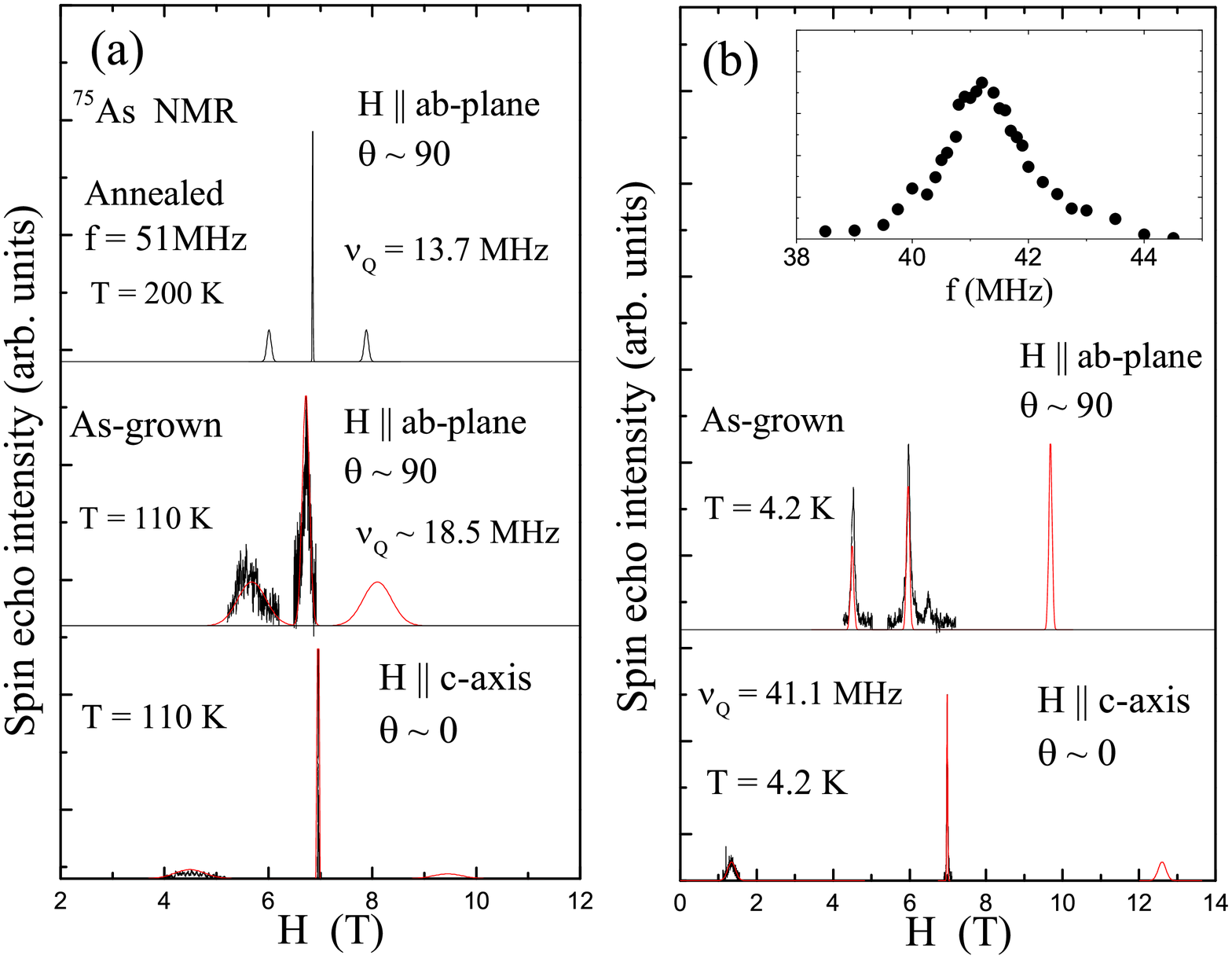} 
\caption{(Color online) (a) Field-swept $^{75}$As-NMR spectra for the as-grown CaFe$_2$As$_2$ crystal (quenched from 960~$^{\circ}$C) 
at $f$ = 51 MHz in the high temperature tetragonal phase (measured at $T$ = 110 K)  for magnetic field parallel to the $c$-axis  (bottom) and perpendicular to the $c$-axis (middle),  together with $^{75}$As-NMR spectrum at $T$ = 200 K  for the annealed CaFe$_2$As$_2$ crystal (quenched from 400~$^{\circ}$C after annealed, see text) with $H$ perpendicular to the $c$-axis (top).
    The black and red lines are observed and simulated spectra, respectively.  
     Expected lines above 8.5 T are not measured due to the limited maximum magnetic field for our SC magnet.
(b) Same as described in (a) but in the low temperature collapsed tetragonal phase (measured at $T$ = 4.2K).  
Inset: $^{75}$As NQR spectrum at $T$ = 4.2 K and $H$ = 0 T.
 }
\label{fig:As-spectrum}
\end{figure}

    Figure \ \ref{fig:As-spectrum}(a)  shows typical field-swept $^{75}$As-NMR spectra of the as-grown CaFe$_2$As$_2$ crystal in the HT ${\cal T}$ phase (measured at  $T$ = 110 K)   for two magnetic field directions of $H$ $\parallel$ $c$-axis  and $H$ $\parallel$ $ab$-plane. 
    The spectra exhibit a typical feature of a nuclear spin $I$ = 3/2 with Zeeman and quadrupolar interactions, which result in a sharp central  transition and two satellite peaks split by the quadrupolar interaction of the As nucleus with the local electric field gradient (EFG). 
   The observed quadrupole-split NMR spectra were reproduced by a simple nuclear spin Hamiltonian 
${\cal H}$ = ${\cal -}$$\gamma$$\hbar$${\vec I}$$\cdot$${\vec H_{\rm eff}}$ +$\frac{h \nu_{\rm Q}}{6}$ [3$I_{\rm z}^{2}$-$I(I+1)$],
where $H_{\rm eff}$ is the effective field at the As site (summation of external field $H$ and 
the hyperfine field $H_{\rm hf}$), $h$ is Planck's constant, and $\nu_{\rm Q}$ is nuclear quadrupole frequency
defined by $\nu_{\rm{Q}}$ =  $eQV_{ZZ}$/2$h$ where $Q$ is the quadrupole moment of the As nucleus, $V_{ZZ}$ is the EFG at the As site. 
   The red lines in the figure show simulated spectra calculated from the simple Hamiltonian with $\nu_{\rm{Q}}$ = 18.5 MHz.  
      It is noted that the satellite linewidth, which reflects the distribution of EFG due to defects or lattice distortion, is relatively large. 
      To reproduce the linewidth, one needs to introduce $\sim$ 15 $\%$ distribution ($\Delta \nu_{\rm Q}$ = 2.7 MHz)  at $T$ = 110 K as shown by red curves in the figure. 
       This is much larger than $\sim$ 4 $\%$ distribution of  $\nu_{\rm Q}$ = 13.7 MHz at $T$ = 200 K for the case of $^{75}$As-NMR in the annealed CaFe$_2$As$_2$ crystal shown at the top panel in  Fig. \ \ref{fig:As-spectrum}(a), which has been reported previously.\cite{Ran2011}
      This indicates  that local As environment in the as-grown CaFe$_2$As$_2$ has a higher degree of inhomogeneity than in the annealed CaFe$_2$As$_2$, which supports the idea that the higher temperature quenching introduces strains inside the crystal.

\begin{figure}[tb]
\includegraphics[width=8.5cm]{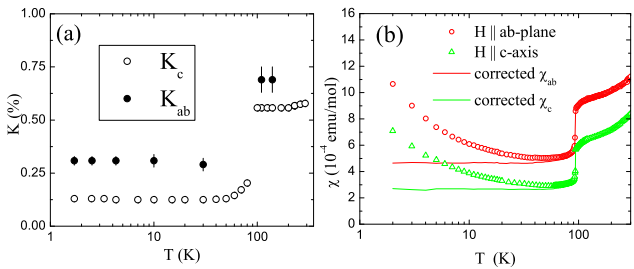} 
\caption{(Color online) (a) Temperature $T$ dependence of $^{75}$As NMR shifts $K_{\rm ab}$ and $K_{\rm c}$ 
for the as-grown CaFe$_2$As$_2$. 
(b) Anisotropic magnetic susceptibility $\chi$ $\equiv$ $M/H$ versus$T$ (where $M$ is magnetization and $H$ is applied magnetic field) for the as-grown CaFe$_2$As$_2$ crystal measured at $H$ = 1 T. 
    From the NMR Knight shift measurements shown in (a), the upturns in $\chi(T)$ below $\sim$ 50 K are not intrinsic, originating from impurities. 
     The solid lines are corrected $\chi(T)$ by subtracting the impurity contributions. 
}
\label{fig:T-K}
\end{figure}

    Below  $T_{\rm s}$ $\sim$ 96 K, the spectra for the as-grown CaFe$_2$As$_2$ crystal  for both $H$ directions change drastically (see, Fig. \ \ref{fig:As-spectrum}(b)).
    This is not due to magnetic ordering but instead due to a dramatic change in $\nu_{\rm Q}$ from  $\sim$18.5 MHz in the HT  ${\cal T}$ phase to  41.1 MHz at $T$ = 4.2 K in the LT c${\cal T}$ phase. 
   The principle axis of the EFG at the As site in the c${\cal T}$  phase is found to be along the crystal $c$-axis, as in the case of the ${\cal T}$ phase, as expected,   since the As site has a local fourfold symmetry around the $c$-axis. 
    The value of  $\nu_{\rm Q}$ = 41.1 MHz is confirmed by the observation of NQR spectrum at zero magnetic field at $T$ = 4.2 K shown in the inset of Fig. \ \ref{fig:As-spectrum}(b), which can be compared with $\nu_{\rm Q}$  $\sim$ 35.8 and 41.5 MHz in the LT c${\cal T}$ phase in (Ca$_{1-x}$Pr$_x$)Fe$_2$As$_2$ for $x$ = 0.075 and 0.15, respectively.\cite{Ma2013} 
    The large change in $\nu_{\rm Q}$ has been attributed to a structural phase transition without any magnetic phase transition.\cite{Ran2011,Ma2013} 
    In addition, we do not see any magnetic broadening in the $^{75}$As NMR spectra even at the lowest temperature 1.5 K  for our measurements. 
      The  $\nu_{\rm{Q}}$ is nearly independent of $T$ below $T_{\rm s}$. 
     This behavior contrasts  with that observed in the HT ${\cal T}$ phase where $\nu_{\rm{Q}}$ decreases from $\sim$ 18.5  MHz at $T$ = 110 K to $\sim$ 18.0 MHz at $T$ = 140 K\cite{Ran2011}, as well as the case in the HT ${\cal T}$ phase of the Sn-flux  CaFe$_2$As$_2$ where  $\nu_{\rm{Q}}$ decreases from $\sim$ 14 MHz at 170 K to $\sim$ 12 MHz at 270 K.\cite{BaekCaFe2As2}
    Since $V_{ZZ}$ arises from hybridization between the As-4$p$ and Fe-3$d$ orbitals with an additional contribution from the non-cubic part of the spatial distribution of surrounding ions, the larger $\nu_{\rm{Q}}$ in  the c${\cal T}$ phase indicates a strong hybridization between the orbitals.
    The difference of $\nu_{\rm Q}$ in magnitude and in its $T$ dependence can be qualitatively interpreted when one takes into consideration the difference in magnitude and $T$ dependence of $c$-axis lattice constant in the corresponding phases:  a nearly $T$-independent behavior ($\sim$ 10.65 $\AA$)  below $T_{\rm s}$ and  a monotonic increase from 11.2 $\AA$ at $T$ = 100 K to 11.58 $\AA$ at 300 K in the ${\cal T}$ phase.\cite{Ran2011}

   Figure \ \ref{fig:T-K}(a)  shows the $T$ dependence of Knight shift, $K_{\rm ab}$ and $K_{\rm c}$ for $H$  
parallel to the $ab$ plane and to the $c$-axis, respectively, where the second order quadrupole shift was corrected for in  $K_{\rm ab}$.
   With decreasing $T$, $K_{\rm c}$ decreases slightly down to $\sim $ 100 K, and shows a sudden decrease at $T_{\rm s}$ similar to the $\chi(T)$ behavior shown in Fig. \ \ref{fig:T-K}(b), and then levels off at low temperatures without showing upturns.
     It is noted that $K_c$ $\sim$ 0.55 $\%$ in the HT ${\cal T}$ phase is greater than $K_c$ = 0.15 -- 0.3 $\%$ for the annealed  CaFe$_2$As$_2$ (not shown here) and Sn-flux  CaFe$_2$As$_2$,\cite{BaekCaFe2As2} but close to $K_c$  = 0.58 $\%$ in Pr-doped CaFe$_2$As$_2$.\cite{Ma2013}  
   $K_{\rm ab}$ also shows the similar $T$-independent behavior in the LT c${\cal T}$ phase: the data are limited above $T_{\rm s}$ due to very poor signal intensity.
    The upturns in $\chi(T)$ observed  at low $T$ in Fig. \ \ref{fig:T-K}(b) are therefore not intrinsic and evidently arise from a small amount of a paramagnetic impurity. 
     The solid lines in Fig. \ \ref{fig:T-K}(b) are corrected $\chi(T)$ obtained by subtracting the impurity contributions, where the $T$-independent $\chi$ indicates a Pauli paramagnetic state, including diamagnetic conduction electron Landau and core-electrons susceptibilities,  for the LT c${\cal T}$ phase.

\begin{figure}[tb]
 \includegraphics[width=5.0cm]{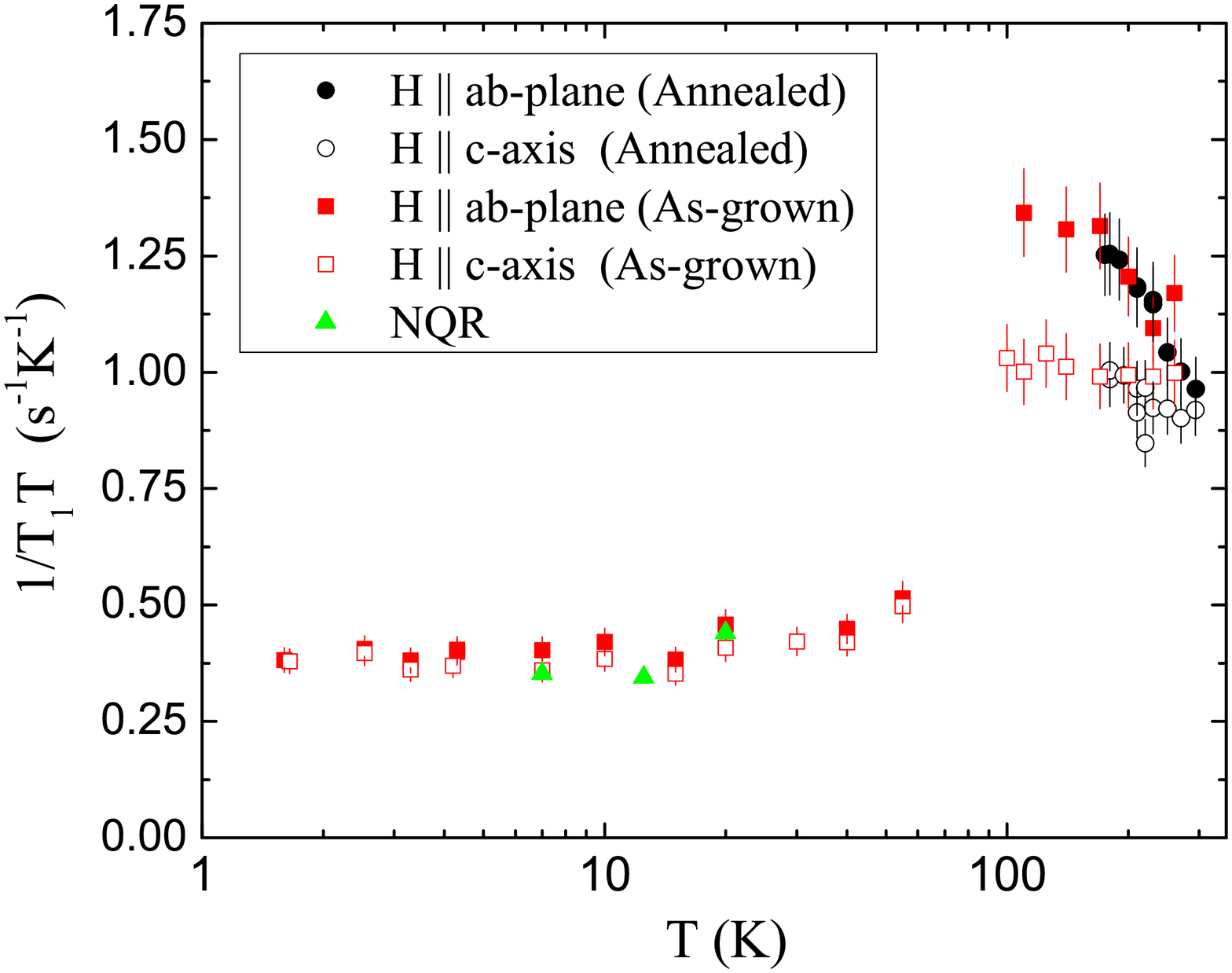} 
 \caption{(Color online) Temperature dependence of 1/$T_1T$ for both magnetic field directions,  $H$ $\parallel$ $c$-axis and $H$ $\parallel$ $ab$-plane and at zero field (NQR) for the as-grown CaFe$_2$As$_2$,  together with the data measured in the annealed CaFe$_2$As$_2$. }
 \label{fig:T1}
 \end{figure}

    Figure \ \ref{fig:T1}  shows 1/$T_1T$ versus $T$ for $H$ parallel and perpendicular to 
the $c$-axis at $H$ = 6 -- 7 T.      
      Above $T_{\rm s}$,    1/$T_{\rm 1}T$  for $H$ $\parallel$ $ab$ plane shows a monotonic increase with decreasing $T$, while  1/$T_{\rm 1}T$ for $H$ $\parallel$ $c$-axis is nearly independent of $T$, similar to the case of the annealed CaFe$_2$As$_2$.   
      The  $T$ dependences of 1/$T_{\rm 1}T$ in the annealed CaFe$_2$As$_2$ are in good agreement with that of 1/$T_{\rm 1}T$ in the Sn-flux CaFe$_2$As$_2$.\cite{BaekCaFe2As2}  
      Below  $T_{\rm s}$, both  1/$T_{\rm 1}T$ decrease suddenly and show Korringa relation $(T_1T)^{-1}$ = constant at low $T$ with no anisotropy in $1/T_1$. 
     1/$T_1T$ is also measured by the $^{75}$As NQR, which are shown in Fig. \ \ref{fig:T1} by solid triangles. 
      The results are in good agreement with the $T$ dependence of  $^{75}$As NQR in the LT c${\cal T}$ phase induced by the application of high pressure 10.8 kbar on CaFe$_2$As$_2$ crystals.\cite{Kawasaki2010, Kawasaki2011} 
     Since the EFG at the $^{75}$As is parallel to the $c$-axis which corresponds to the quantization axis, the relaxation is expected to be the same with that obtained from the NMR with $H$ $\parallel$ $c$ axis. 
    Actually, the $1/T_1T$ data by NQR at zero field are in good agreement with that of NMR with $H$ $\parallel$ $c$ axis, indicating the absence of magnetic field effects on $T_1$ values.

\begin{figure}[tb]
\includegraphics[width=8.8cm]{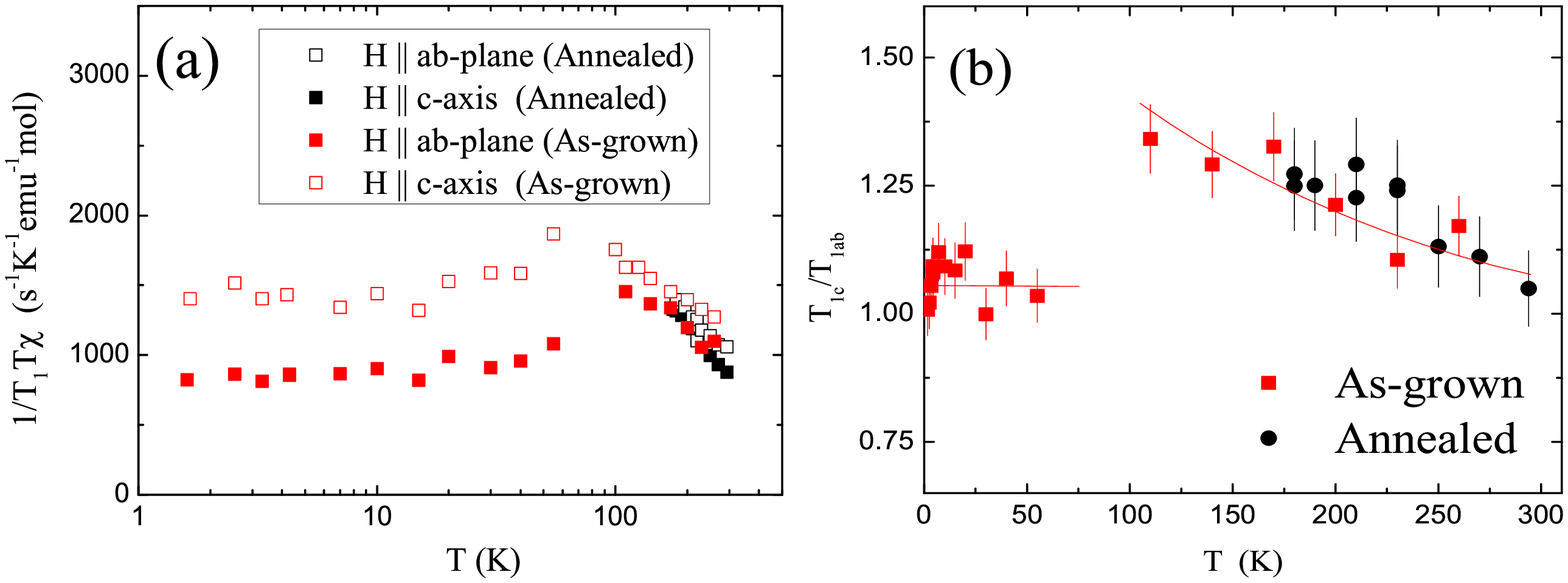} 
\caption{(Color online) (a) 1/$T_1T$$\chi$  versus $T$ for both magnetic field directions, 
$H$ $\parallel$ $c$-axis and $H$ $\parallel$ $ab$-plane in the as-grown sample. 
   $T$ dependences of 1/$T_1T$$\chi$ for both $H$ directions in the annealed CaFe$_2$As$_2$ are also plotted for comparison.
   The increase of  1/$T_1T$$\chi$ observed above $T_{\rm s}$ indicates the growth of the strip-type AFM spin correlations.   
  (b) $T$ dependence of the ratio $r$ $\equiv$ $T_{\rm 1,c}$/$T_{\rm 1,ab}$.   
   The solid line is an eyeguide.  }
\label{fig:T1Tchi}
\end{figure}

   In order to see spin fluctuation effects in the paramagnetic state, it is useful to re-plot the data by changing the vertical axis from $1/T_1T$ to $1/T_1T\chi$ as shown in Fig.~\ref{fig:T1Tchi}, where the corresponding corrected $\chi$ was used for each $H$ direction. 
    $1/T_{\rm 1}T$ can be expressed in terms of the imaginary part of the dynamic susceptibility $\chi^{\prime\prime}(\vec{q}, \omega_0)$ per mole of electronic spins as,\cite{Moriya1963}
$\frac{1}{T_1T}=\frac{2\gamma^{2}_{N}k_{\rm B}}{N_{\rm A}^{2}}\sum_{\vec{q}}|A(\vec{q})|^2\frac{\chi^{\prime\prime}(\vec{q}, \omega_0)}{\omega_0}$, 
where the sum is over the wave vectors $\vec{q}$ within the first Brillouin zone, $A(\vec{q})$ is the form factor of the hyperfine interactions and $\chi^{\prime\prime}(\vec{q}, \omega_0)$  is the imaginary part of the dynamic susceptibility at the Larmor frequency $\omega_0$.  
    On the other hand,  the uniform $\chi$ corresponds to the real component 
 $\chi^{\prime}(\vec{q}, \omega_0)$ with $q$ = 0 and $\omega_0$ = 0. 
    Thus a plot of $1/T_{\rm 1}T\chi$ versus $T$ shows the $T$ dependence of  $\sum_{\vec{q}}|A(\vec{q})|^2\chi^{\prime\prime}(\vec{q}, \omega_0)$ with respect to that of the uniform susceptibility $\chi^{\prime}$(0, 0). 
    Above $T_{\rm s}$, $1/T_{\rm 1}T\chi$ for $H$ $\parallel$ $c$-axis and $H$ $\parallel$ $ab$-plane in both samples increase with decreasing temperature. 
    The increase implies $\sum_{\vec{q}}|A(\vec{q})|^2\chi^{\prime\prime}(\vec{q}, \omega_0)$ increases  more than $\chi^{\prime}$(0, 0), which is due to a growth of spin correlations with $q$ $\neq$ 0, stripe-type AFM wave vector $q$ = $Q_{\rm AF}$ as discussed in the following. 
    Thus we conclude that strong AFM spin fluctuations are realized in the HT ${\cal T}$ phase in both the annealed and as-grown CaFe$_2$As$_2$ crystals, consistent with INS measurements.\cite{Soh2013,Pratt2009}
    It should be noted that nearly $T$-independent behavior of $1/T_1T$ does not always indicate the absence of the AFM spin correlations, as has been observed for $H$ $\parallel$ $c$-axis. 
    One should compare the $T$ dependence of 1/$T_1T$ with that of the uniform susceptibility.
    In the LT c${\cal T}$ phase, on the other hand, $1/T_{\rm 1}T\chi$ are nearly independent of $T$  showing that the $T$ dependence of $\sum_{\vec{q}}|A(\vec{q})|^2\chi^{\prime\prime}(\vec{q}, \omega_0)$ scales to that of $\chi^{\prime}$(0, 0).
    This indicates no significant effects of the AFM spin correlations in the LT c${\cal T}$ phase.

   Now, based on these $T_1$ results, we discuss more details of Fe spin fluctuations in the HT ${\cal  T}$ and LT c${\cal T}$ phases. 
   According to previous NMR studies performed on Fe pnictides,\cite{KitagawaSrFe2As2,SKitagawaAF-Fluctuation,FukazawaKBaFe2As2} and SrCo$_2$As$_2$,\cite{PandeySrCo2As2}  
   the  ratio $r$ $\equiv$ $T_{1,c}$/$T_{1, ab}$ depends on AFM spin correlation modes as 
\begin{eqnarray}
r =  \left\{
\begin {array}{ll}

0.5 + \left(\frac{S_{ab}}{S_c}\right) ^2    \mbox{~ for  the stripe AFM fluctuations} \\ 

0.5     \mbox{~ for  the N\'eel-type spin fluctuations} \\
\end {array}
\right .
\label{eqn:correlation}
\end{eqnarray}
where  ${\cal S}_{\alpha}$ is the amplitude of the spin fluctuation spectral density at NMR frequency along the $\alpha$ direction. 

    As plotted in Fig.\ \ref{fig:T1Tchi}(b), the $r$ is greater than unity and increases with decreasing $T$ in the HT ${\cal T}$ phase for both the as-grown and the annealed crystals.
    Together with the increase of $1/T_1T\chi$ shown in Fig. \ \ref{fig:T1Tchi}(a), we conclude that stripe-type AFM spin fluctuations are realized in the HT ${\cal T}$ paramagnetic state.
    N\'eel-type spin fluctuations can be clearly ruled out because according to Eq. \ \ref{eqn:correlation} that would require $r$  = 0.5, which is in conflict with our measurements shown in Fig.\ \ref{fig:T1}(b) that gives $r$ $>$ 1.0.

      In contrast, the $r$ is close to unity and is $T$-independent in the LT c${\cal T}$ paramagnetic state. 
     Given the fact that we do not detect any trace of the AFM spin correlations in the $T$ dependences of $1/T_1T\chi$ and NMR spectrum, we conclude that electron correlations disappear in the phase. 
     This is consistent with  inelastic neutron scattering measurements\cite{Soh2013,Pratt2009} which demonstrate the evidence of the absence of magnetic fluctuations in the non-superconducting c${\cal T}$  phase in CaFe$_2$As$_2$. 
     According to recent ARPES measurements on CaFe$_2$As$_2$,\cite{Dhaka2014} the multi-band structure can be seen in the HT ${\cal T}$ phase, while the hole pockets around ${\it \Gamma}$ point sink below Fermi energy, resulting in losing the multi-band nature in the LT c${\cal T}$ phase.
   Since the stripe-type AFM spin correlations originate from the interband correlations, the absence of the stripe-type AFM spin correlations in the LT c${\cal T}$ phase is also consistent with the ARPES measurements.

     In summary, we report $^{75}$As NMR and NQR results on the LT c${\cal T}$ and HT T phases in the as-grown CaFe$_2$As$_2$. 
  $^{75}$As NMR and NQR spectra measurements confirm no static magnetic ordering in the LT c${\cal T}$ phase.
     Through the $T$ dependence of 1/$T_1$, Knight shift and magnetic susceptibility $\chi$,  stripe-type AF spin correlations are realized in the HT ${\cal T}$ phase in both the as-grown and annealed CaFe$_2$As$_2$ crystals.
   On the other hand, no trace of the dynamical AFM spin correlations can be found in the non-superconducting LT c${\cal T}$ phase. 
   The lack of any magnetic broadening of NMR spectrum and $T$-independent Knight shift demonstrate no  development of static Fe spin correlations in the c${\cal T}$ phase.       
   These observations, combined with the recent INS measurements showing the absence of magnetic fluctuations, bring us to  the conclusion that electron correlations completely disappear in a wide energy scale from NMR to INS techniques in the  non-superconducting  c${\cal T}$ phase in CaFe$_2$As$_2$. 
    These results are consistent with the general view that spin correlations play an important role for the appearance of unconventional SC in the iron pnictides.
     In this point of view, we speculate that SC observed in the c${\cal T}$ phase of (Ca$_{1-x}$Sr$_x$))Fe$_2$As$_2$ and (Ca$_{1-x}$R$_x$)Fe$_2$As$_2$ (R = Pr, Nd) is probably not associated with the c${\cal T}$ phase but with some other phase. 
   Further detailed studies to elucidate the source of superconductivity in (Ca$_{1-x}$Sr$_x$))Fe$_2$As$_2$ and (Ca$_{1-x}$R$_x$)Fe$_2$As$_2$ are highly required.

The research was supported by the U.S. Department of Energy, Office of Basic Energy Sciences, Division of Materials Sciences and Engineering. Ames Laboratory is operated for the U.S. Department of Energy by Iowa State University under Contract No.~DE-AC02-07CH11358.

\end{document}